# Effects of Mobile Gacha Games on Gambling Behavior and Psychological Health

Ji Woo Han

**[Abstract]**

The study examines how Gacha games affect gambling practices by analyzing when players start playing and how often they play and how much money they spend on games. The modified Problem Gambling Severity Index (PGSI) serves as the research tool to examine the impact of these variables on mobile Gacha game gambling-like severity. A survey distributed to online Gacha game communities provided the data which researchers analyzed by applying linear regression and multiple regression and T-test methods. The research indicates entry age demonstrated low association with gambling severity but gameplay frequency together with duration directly influenced higher gambling severity ratings. The research findings demonstrated that the level of Gacha game spending directly influences gambling severity scores among players. The research findings demonstrate how addictive potential exists in mobile Gacha games which requires stronger regulation of these platforms for young vulnerable gamers. Research needs to study the extended influence of Gacha game exposure while examining how cognitive biases affect gambling behavior patterns.

## *[1. Introduction]*

### *1.1 Background & Motivation*

The increasing prevalence of mobile gacha games that are linked to elements of chance and in-game purchases have been discussed with their influence on gambling and psychological health (Lakić et al., 2023).[1] Despite the abundance of studies dedicated to the impact of traditional gambling on psychological functioning (Grant et al., 2016; Livazović & Bojčić, 2019; Pietrzak & Petry, 2006)[234], the correlation between gacha games' mechanics and gambling-like behaviors has not been investigated thoroughly. Gacha games share significant similarities with gambling, as they also involve the use of chances for bonuses and the necessity to spend money for desired results (Rogers, 1998).[5] These similarities imply that mobile gacha games could foster behaviors related to gambling addiction and potentially contribute to gambling addiction, particularly due to their widespread accessibility, which could easily influence adolescents and young people and lead to negative future outcomes (Emond et al., 2019).[6]

### *1.2 Current Problems*

One of the primary concerns is the insufficient regulation of age restrictions for mobile gacha games. Unlike traditional forms of gambling that are strictly regulated and accessible only to individuals over 18, gacha games often have lower age ratings, such as 12+ for popular titles like *Genshin Impact*. This lack of stringent oversight allows vulnerable adolescents to engage

---

with these games. Therefore, this paper examines how the age at which individuals begin playing gacha games and prolonged exposure to these mechanics influence the severity of gambling behavior. Using a quantitative research method through surveys, we explore the correlation between early exposure and the development of gambling tendencies. This accessibility heightens the risk of desensitization to gambling or the development of gambling disorders.

*1.3 Importance & Significance*

        The significance of this research helps businesses and governments develop effective health promotion policies and practices, and foster education initiatives. Mobile gacha games are designed in a way that resembles gambling, and in fact, the random reward mechanisms are very similar to gambling, and the evidence suggests that they can cause gambling disorders, especially in susceptible groups such as adolescents (Kamamura et al., 2022).[7] The randomness of the gacha games has been said to give similar effects to people as those of gambling, for example the need to spend more time and money in order to achieve the wanted results. Since these games could create addictive tendencies and lead to financial losses, it is important to look at the negative effects of the games fully. Therefore, concentrating on the social evils of mobile gacha games on the players of the games, this paper aims at exposing the risks of such games. The findings can help policymakers identify the necessity of regulation, assist public health professionals in developing prevention and treatment strategies, and provide educators and parents with the knowledge necessary to guide young people away from potentially harmful tendencies in gaming habits. Therefore, this research aims to focus on the potential risks of mobile gacha games and to call for action to prevent their impact on society.

---

[7] Kamamura, T., Koyama, Y., Mori, T., Motonishi, T., & Ogawa, K. (2022). Loot box gambling and economic preferences: a survey analysis of Japanese adolescents and young adults. *Applied Economics*, *55*(44), 5213–5229. https://doi.org/10.1080/00036846.2022.2138817

*[2. Literature Review]*

*2.1 Summary of Current Research Trends*

Mobile gacha games have significantly impacted the gaming industry and the study of gambling behaviors in contemporary society. Starting from the first gacha game, Dragon Collection by Konami in 2010, gacha games have developed into complex mechanisms that heavily influence players' actions and have garnered much interest from both businesses and scholars (Rentia & Karaseva, 2022).[8] Nevertheless, the connection between mobile gacha game mechanics and gambling-like behaviors has not yet been sufficiently researched, which reveals several significant gaps in the existing literature.

**Development of Mobile Gacha Games and Early Research**

Mobile gacha games have rapidly gained popularity, particularly after 2015 when the game designs became more advanced. These designs included features that increased the user's level of interaction and spending, for instance, "limited-time" events that make players spend more time and money (Shibuya et al., 2022).[9] The global success of such games as Genshin The success of Genshin Impact in 2020, generating $1.5 billion in revenue and attracting 7.9 million users at its peak, underscores the widespread appeal of the gacha model (Wylie, 2024).[10] Notably, this was just the beginning; the game continued to grow, reaching over 139 million downloads by the end of 2022 and surpassing $4 billion in total revenue, demonstrating the sustained popularity and profitability of gacha mechanics. However, the early studies on the risk of developing mobile gacha games were scarce and based on the qualitative comparison of gacha mechanics with traditional gambling without supporting quantitative data.

*2.2 Defined Applicable Theories*

**Problem Gambling Severity Index**

In the case of mobile gacha games, the PGSI is regarded as an especially important assessment tool. The Problem Gambling Severity Index (PGSI) is one of the instruments that are included in the Canadian Problem Gambling Index (CPGI) created in Canada (Ferris et al. , 2011).[11] The PGSI is a nine-item scale which assesses the different aspects of gambling problems such as the frequency of gambling, the monetary impact and psychological distress (Tang et al. , 2022).[12] It is meant to assess the extent of gambling behaviors and the effects that are related to it. Respondents rate each item using the frequency scale of 0 (never) to 3 (almost always); higher scores indicate higher risk. The total score is an efficient tool for definition and measurement of gambling related issues as it separates people based on their scores into groups starting from those with no gambling issues and ending with those experiencing pathological gambling.

Mobile gacha games often include aspects that are similar to gambling, and this can contribute to the formation of gambling-like behaviors, especially among the young persons (Kamamura et al. , 2022).[13] In this research, the PGSI is used as the dependent variable to measure how playing mobile gacha games is associated with gambling severity. In particular, this research examines three hypotheses concerning the effect of gacha game engagement on gambling behaviors, as measured by the PGSI tool. The categorical organization of the PGSI makes it possible to examine patterns like the escalation of spending and the repeated use of gacha mechanics that may lead to more severe gambling problems in the future (Ferris et al. ,

---

2001; Tang et al. , 2022).[14][15] With Problem Gambling Severity Index, this research will systematically examine the impacts of mobile gacha games on gambling behavior and psychological well-being and possible risks related to such games.

**Lottery Theory**

Mobile gacha games which depend on probability mechanics similar to lotteries are based on several psychological factors that can affect players' behavior. An example of such a bias is the 'Illusion of Control', in which players are duped into thinking that the choices they make in the game have an impact on events that are entirely random (Rogers, 1998).[16] This mirrors behaviors observed in lottery participants, who often select numbers or employ rituals they believe will increase their chances of winning. Additionally, the 'Sunk Cost Fallacy' is prevalent in gacha games, where players continue to invest time and money due to previous expenditures, despite the low probability of significant returns (Rogers, 1998).[17] This cognitive entrapment reinforces the cycle of spending, as players become increasingly committed to the game in hopes of obtaining rare rewards.

Furthermore, the 'Near Miss' effect plays a crucial role in maintaining player engagement in mobile gacha games (Rogers, 1998).[18] When players perceive that they almost obtained a rare item, they are more likely to continue playing, driven by the belief that success is just around the corner. This effect is compounded by social factors, such as seeing others' successes or being part of a community that encourages continued participation (Rogers, 1998).[19] These elements based

---

on the psychology of the lottery gambling indicate that gacha games have the prospects of promoting gambling-like behavior and adolescents and young adults whose psychological mechanisms are not fully developed (Rogers, 1998).[20] It is crucial to comprehend these psychological processes to evaluate mobile gacha games' effect on gambling and psychological health.

**Developmental & Behavioral Psychology**

Young people are the most vulnerable to the reward system of mobile gacha games because of their vulnerabilities due to insufficient development in their cognitive functions (Galvan, 2010).[21] Research has established that problem gambling in adolescents is highly linked with unfavorable psychological, social, and financial consequences (Livazović & Bojčić, 2019).[22] Emond and Griffiths (2020) further explored these vulnerabilities, noting that while only a minority of adolescents develop gambling disorders, the increasing prevalence of online gambling-like mechanisms, such as those in mobile gacha games, could exacerbate this issue.[23]

Additionally, Purwaningsih and Nurmala (2021) identified a significant negative correlation between problematic online gaming and adolescent mental health, suggesting that gaming addiction may lead to broader psychological issues associated with gacha games.[24] This underscores the need for research focused on the age-related risks of early exposure to mobile gacha games and its potential long-term impact on gambling tendencies and psychological well-being.

---

*2.3 Research Gap*

However, several gaps have been identified in the existing literature on mobile gacha games that need to be filled. First, the literature review shows that there is a shortage of quantitative studies that would focus on the similarities between gacha games and gambling. Although there are qualitative works explaining that gacha mechanisms are similar to gambling, there is a lack of precise and objective evidence concerning the level of addiction, the frequency of spending, and psychological dependency on gacha games. Further studies should use the PGSI to measure these factors, which would help determine if gacha games are as dangerous or more dangerous than conventional gambling.

Second, the addictive nature of mobile gacha games themselves, independent of game design elements, has not been thoroughly explored. Longitudinal studies are needed to examine whether initial exposure to a single gacha game leads to increased participation in other gacha games or different forms of gambling over time. Such research should also investigate the role of individual differences, such as impulsivity and cognitive biases, in predicting addiction risk.

Third, there is a lack of knowledge regarding the influence of the age when players start playing mobile gacha games on their future gambling habits. Cross-sectional comparisons are essential to determine when people are most susceptible to gacha games' addictive characteristics and whether early exposure to them raises the likelihood of developing gambling disorders in the future.

Finally, while traditional gambling has been extensively studied from a behavioral psychology perspective, similar research on mobile gacha gamers is scarce. Future studies should focus on the psychological triggers and reinforcement mechanisms that sustain gacha game behavior, such as immediate gratification, near-miss scenarios, and social influences.

Understanding these mechanisms is vital for developing targeted intervention strategies and effective regulatory policies.

*[3. Research Framework]*

*3.1 Research Purpose*

From the identified research gaps this study aims to examine the age factor in terms of gambling severity in mobile gacha games by analyzing the correlation between various factors such as gacha game playing frequency, duration, intention to pay and money spent by South Korean gacha game players. Furthermore, this study will also be investigating the correlation between gambling severity in gacha games and the psychological well-being along with addictiveness of the gacha itself to briefly spark the discussion. Our goal is to ultimately uncover if starting gacha and playing gacha games earlier in life increases the likelihood of having severe gambling issues. However, due to research done on mobile gacha games is relatively deficient, the study also aims to spark discussion in multiple directions regarding the research topic.

With the discoveries made through the research, this study intends to provide insight to government organizations and businesses to consider the impacts that mobile gacha games could pose to customers who are still in the developmental stage and ultimately impose necessary measures to prevent negative impacts to the customers and the society as a whole.

*3.2 Research Question*

1. How does the age of getting involved in gacha games and prolonged exposure to such game designs influence the severity of the gambling behavior?

2. How does frequency of playing gacha games and being involved in gacha affect decision-making skills in terms of finance and other psychological well-being of a person?

*3.3 Research Model*

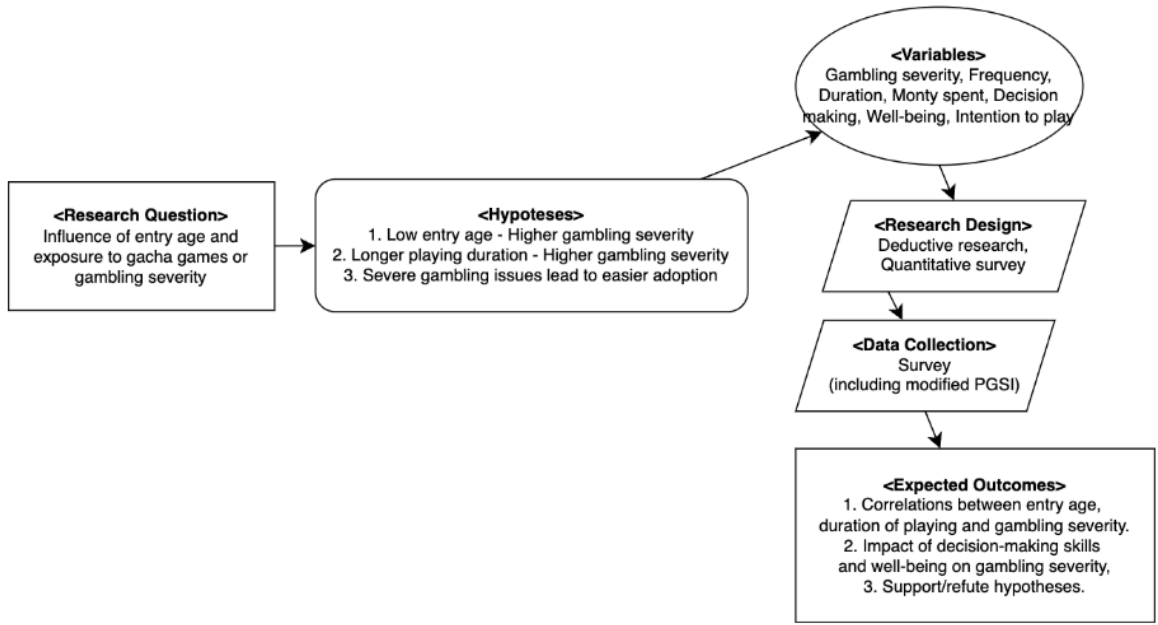

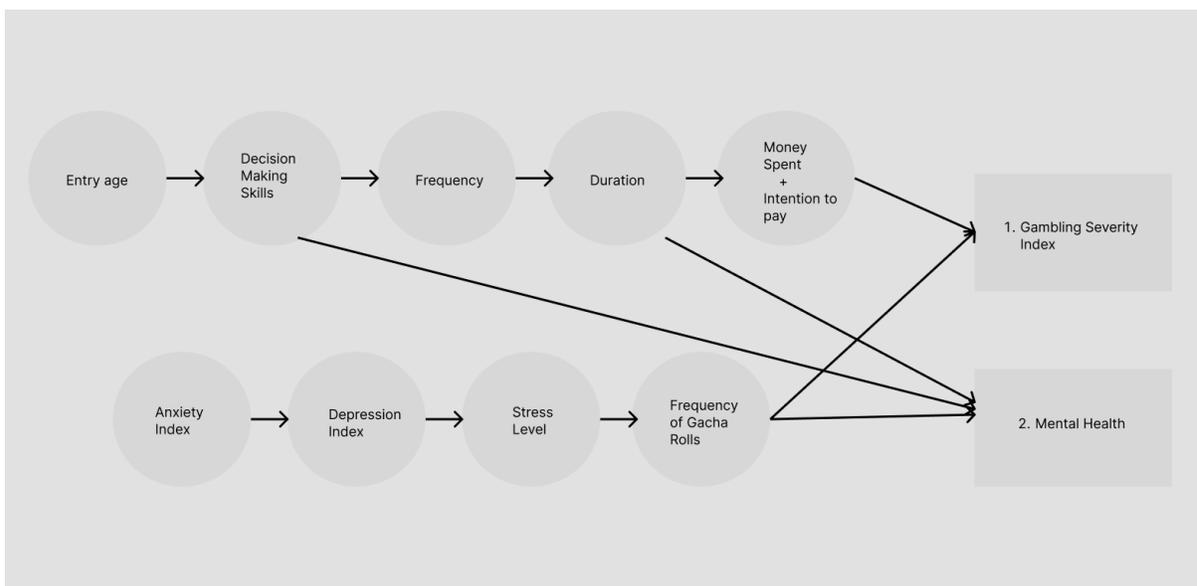

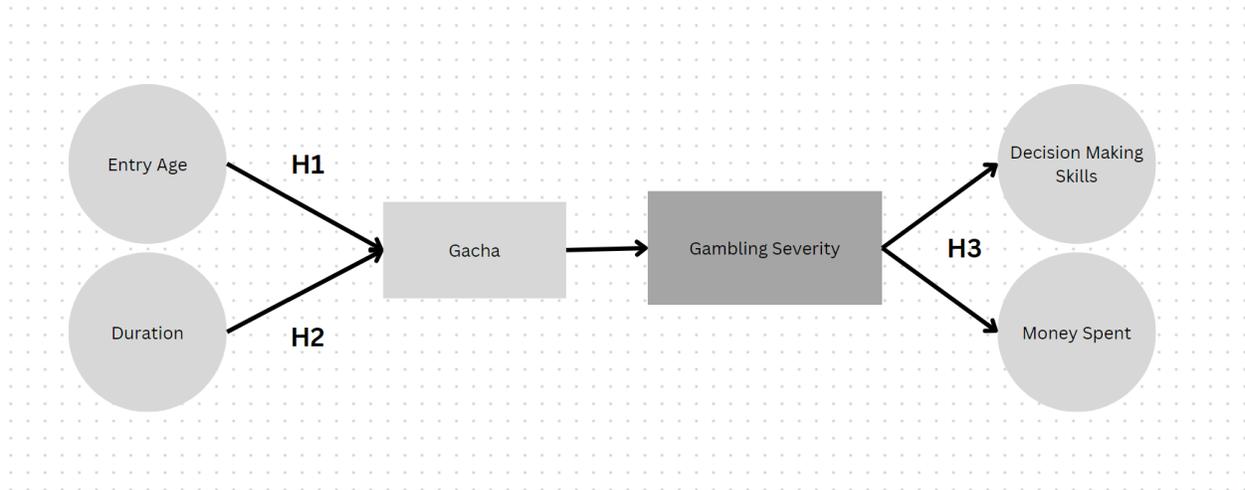

*3.4 Hypothesis & Variables*

**H1: The lower the entry age is, the higher/more severity of the gambling behavior.**

The first hypothesis will test the correlation between the entry age of playing mobile gacha games and being involved with gacha and the severity level of the gambling behavior. Since adolescents and young children are substantially vulnerable to gambling influences due to cognitive and developmental immaturities[25], this study will assume that such theory would apply the same with gacha games. Thus, the age factor as a variable will play an important role in finding the correlation between entry age and gambling severity level.

**H2: The longer people play the Gacha game, the higher/more severity of the gambling behavior.**

The second hypothesis will examine the relationship between the duration of playing mobile gacha games and the severity of the gambling behavior. In traditional gambling, it has been observed that the longer a gambler's duration in gambling, the worse the problem gambling

tendencies becomes.[26]

**H3: Once someone reaches a severe level of gambling problem, it becomes much easier for them to join other gacha games and spend money. (people play more gacha games and spend more money)**

The last hypothesis assumes that once a user reaches a severe level of problem gambling, the user becomes addicted to the gacha itself and is more likely to be involved in multiple mobile gacha games because the user is influenced by the gacha system itself more than the game design itself. The assumption is backed by the theory of gambling psychology where problem gamblers' brain activation is associated with the gambling game's reward rather than the gamble itself.[27] Through the observation of this hypothesis, the study could get one step closer to uncovering the potency of mobile gacha games and the potential risk factors associated with playing gacha.

**Variables:**

1. Gacha Severity

   a. Measured dependent variable with the gambling severity index to determine a person's problem gambling severity level. This index will be cross-examined with the other independent variables to determine the correlation.

      i. *It is measured through our modified PGSI questionnaire (Link:*
         *https://forms.gle/SXAQMU5CZVhqUm7J9)*

2. Entry Age:

   a. It is one of the independent variables, which measures the age at which people started to play any gacha games.

   b. Measured by survey questionnaire:

---

      i.     *1. How old were you when you first started playing a gacha game?*

3. Frequency of playing gacha games (days per week)

    a. Measured independent variable by the number of days in a week a user plays gacha games.

    b. Measured by survey questionnaire:

      i.     *3. How many days a week do you play gacha in the game?*

4. Duration (in hours per day and years played)

    a. Measured independent variable by the length of a user's involvement in gacha games over their lifetime.

    b. Measured by survey questionnaires:

      i.     *2. Since you first played a gacha game, how many years have you been playing gacha games?*

      ii.    *9. Once you start playing, how many hours do you play on average in 24 hours?*

5. Money Spent

    a. Measured independent variable in Korean Won on how much a user has spent in gacha (per session? Per event?).

    b. Measured by survey questionnaire:

      i.     *5. What was the maximum amount of money you have spent on a gacha each time? (Or if you could describe how much money you've spent during an event or any specified duration of time, please let us know)*

6. Intention to Pay

    a. Measured independent variable in Korean Won on how much a user is willing to spend in gacha rolls per session.

    b. Measured by survey questionnaire:

        i. *4. What is the maximum budget you allow yourself to spend every time you play/roll for gacha?*

7. Financial Decision Making Skills (mention the survey Q: intention to pay vs actual paid/overpaid)

    a. Measured independent variable in the will power of a user to make sound financial decisions or be aware of impulsive purchases.

    b. Measured by the discrepancy between two questionnaires:

        i. *6. Have you ever spent more money than expected or planned to get items or characters you want in the game including the gacha system?*

        ii. *6-1. If you answered "yes" in the above question, how much have you spent?*

        iii. *4. What is the maximum budget you allow yourself to spend every time you play/roll for gacha?*

## *[4. Research Plan]*

To discover the correlation between the severity of mobile Gacha games and psychological effects of them, this overall progress and condition of this study are as follows.

### *4.1 Target Sampling*

The subject of this research is Koreans who have played mobile Gacha games and have spent money in the games. The age group of the study was not specified to extensively analyze Gacha games players' behavioral patterns across various age groups. This is because the Gacha game attracts many users from a wide age range; also, this research aims to examine differences in behavior depending on the game start age and duration of playing them. Some studies show that gambling involvement tends to be noticeable until mid-adolescence, increase to the twenties, and decrease in old age.[28] This phenomenon could be shown in Gacha games. Therefore, it is important to know the effects of the games starting age and duration of playing them. Due to this widespread accessibility of Gacha games, this population is well-suited to offer a comprehensive understanding of the game users in terms of game addiction.

To collect data, this study selected the mobile Gacha game-related community as the main sample population. More specifically, the survey was conducted on the following major Gacha games communities sites: the community of "Genshin" (https://genshin.inven.co.kr/), the community of " Unknown Knight" (https://game.naver.com/lounge/UnknownKnight/home), the mommunity of "Puzzle and Dragon" (https://pad.inven.co.kr/). These communities could be expected to properly reflect the characteristics of the research subject, as users who often play

---

Gacha games and show high interest and involvement in the games.

**Sample size for testing a correlation coefficient using Fisher's z-transformation**

**Input and calculation**

True correlation [0.5]

Correlation under null hypothesis [0]

Alpha [0.05]

Power [0.8]

[Calculate]

The required sample size is 30.

[Copy result statement to clipboard]

**Options**

Calculate
- ⦿ Sample size
- ○ Power (output decimal places: [4])

Advanced
- ☐ Account for drop-outs
- ☐ Bonferroni correction

The real sampling size of this study is 42 people. To evaluate whether this size is sufficient to produce statistically significant results, an online sample size test (https://homepage.univie.ac.at/robin.ristl/samplesize.php?test=correlation) was conducted. The test applied power analysis to consider the condition where there is only one population. Also, correlation test was utilized to calculate the size because hypothesis 1 and 2 using linear regression. The null hypothesis set under the correlation test menu; 1) entry age does not affect gambling problem severity 2) the amount of money spent does not affect gambling problem severity. The research team set the true correlation coefficient as 0.5 and the correlation under null hypothesis as 0. Based on this setting, the required sample size is 30. In other words, the sample size must be at least 30 to have statistically valuable results. Thus, the number of real participants is meaningful enough to verify all of the hypotheses in this study with reliability and validity of the results.

*4.2 Data Collection Methods*

In this study, data is collected through survey questionnaires. As mentioned above, the survey will be distributed preliminary on the online Gacha gaming communities to gather data for users who have spent money playing the games. The sampling method was random, which allows diverse participants to voluntarily respond to the survey over a certain period of time (for 4 days). Every respondent will answer questions about their experiences in the Gacha games such as frequency of the games, money spending patterns on them, and gaming tendency.

The questionnaire is based on the Problem Gambling Severity Index (PGSI). PGSI is a widely-used self-report measure consisting of 9 items for problem gambling behavior in the general population.[29] In 1996, a 31 items measurement tool called CPGI, which stands for Canadian Problem Gambling Index, was created due to a project to develop a new method to estimate problem gambling in Canada.[30] Particularly, among the questions of CPGI, 9 items are used to calculate the prevalence of problem gambling, which is PGSI. This index was modified for situations about the Gacha game in this study. These adjustments could more accurately reflect the user characteristics within the Gacha games to precisely evaluate the subjects of this study.

*4.3 Survey Sample Questions*

1. How old were you when you first started playing gacha games?

2. Since your first gacha game, how many years have you been playing gacha games?

3. How many days a week do you play gacha in the game?

- 1~2 days

- 3~4 days

- 5~6 days

- Everyday of a week

4. What was the most you've spent in the gacha

**Modified PGSI: 0 to 3**

1. Have you ever spent on gacha more than you could really afford to lose?

2. Have you needed to gacha with larger amounts of gacha rolls to get the same feeling of excitement as before?

3. When you did gacha and did not get what you wanted, did you ever spend more to compensate for the losses and get the item/character you really wanted?

4. Have you borrowed money or sold anything or used a pre-payment system such as mobile payment or credit payment to do gacha?

5. Have you felt that you might have a problem with doing too much gacha?

6. Has playing gacha ever caused you any health problems, including mental health (ex. Anxiety and stress)?

7. Have people criticized you gacha or told you that you had a gambling problem, regardless of whether you thought it was true?

8. Has your involvement in gacha caused any financial problems for you or your household?

9. Have you felt guilty about the way you play gacha or what happens when you play gacha?

*4.3 Data Analysis Methods*

## Results

|  | | Score (0-27) | Severity Index |
|---|---|---|---|
|  | Total | 18 | Problematic |

For ease of reference, a severity index is provided for the total score based upon prior research. Scores of 0 are considered to be a severity of "none", scores of 1-4 are considered "low", scores of 5-7 are considered "moderate" and scores of 8 and above are considered to be "problematic" severity (Currie et al., 2010; Currie et al., 2013).

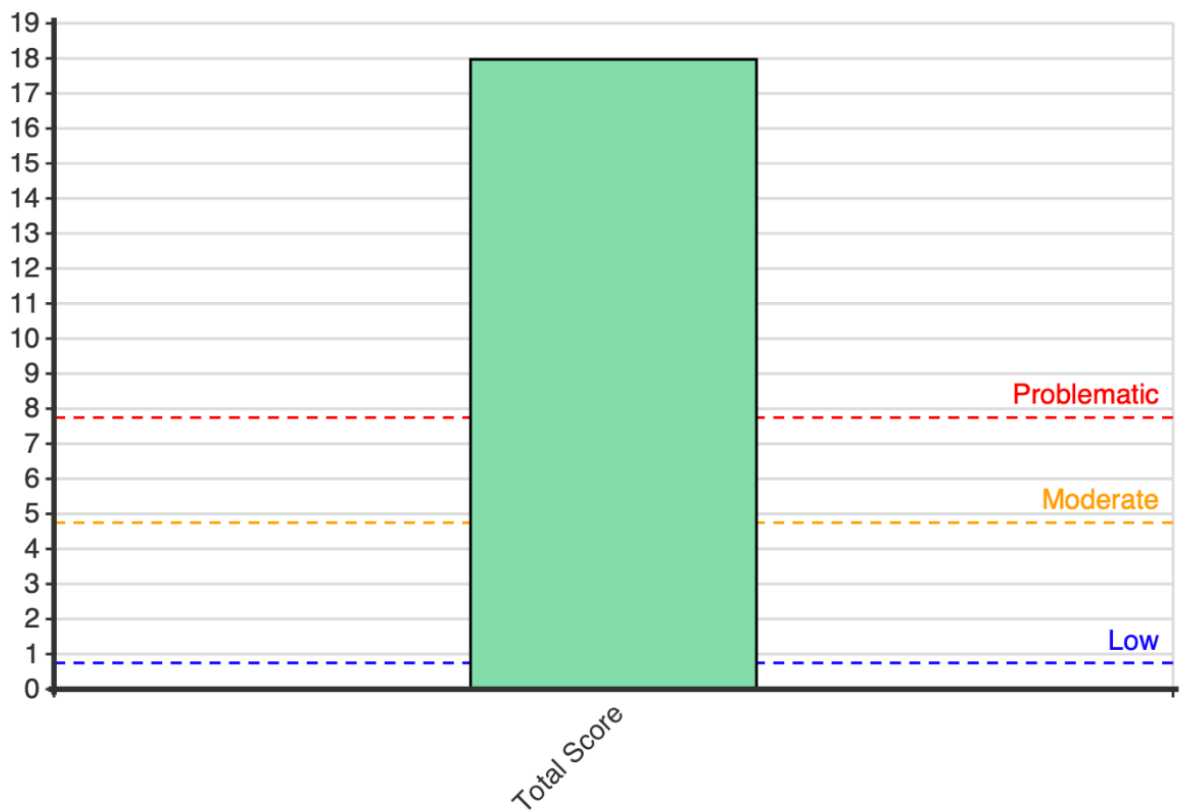

The original *Problem Gambling Severity Index* comes with a user manual that guides you through how to utilize the index tool.[31] The result thresholds can be divided into three major sections which are, non-problem gamblers (PGSI = 0), low risk (PGSI = 1-2), moderate risk

---

(PGSI = 3-7) and problem gamblers (PGSI = 8+).[32] This study will follow the original guidelines even in the modified PGSI model since the intrinsic variables and the theoretical foundations are the same as the original PGSI model. The questions are only modified in the slightest to convert all the "gambling" subjects into the context of "gacha".

Both hypothesis 1 and 2 will be run through linear regression analysis in order to predict the larger population's behavior with respect to the population sample that this research collects. In hypothesis 1, the dependent variable will be the severity of gacha game behavior (*sum of PGSI*) and the independent variable will be the entry age for gacha game (in years). For hypothesis 2, the dependent variable will also be the severity of gacha game behavior (The sum of modified PGSI) and the independent variable will be the duration of playing gacha games (in hours per day). For further analysis, these two hypotheses will also be run through the R program for multiple regression analysis for more accurate results and clear comparison in the two hypotheses.

Finally, hypothesis 3 will be tested utilizing independent T-test to compare the mean of two variables. The two dependent variables will be set as the number of gacha games played (when playing the most amount of games) and the amount of money spent (in Korean Won). These two variables will be tested to analyze the relationship between the severity of gacha game behavior (The sum of modified PGSI).

H1: (Linear regression)

- DV: the severity of gacha game behavior (The sum of modified PGSI)

- IV: the age (in years)

For analysis:

---

- Dichotomous DV - Continuous IV: Logistic regression

- Continuous DV - Continuous IV: Linear regression

H2: (Linear regression)

- DV: the severity of gacha game behavior (The sum of modified PGSI)

- IV: the duration of playing gacha games (hours per day)

H3: independent T-test

- DV1: the number of playing gacha games (when playing the most amount of games)

- DV2: the amount of money spent on gacha games (in Korean Won)

- By groups: 1) low-moderate players, 2) severe players

*4.4 Research Limitations*

One of the main limitations of this study is the sampling method. The biggest challenge is that it is hard to determine the exact population size, which is the number of people who have played Gacha games in Korea. As a sort of a trade secret, accurate data on the number of users of Gacha games is information that can only be accessed internally by game companies. Thus, external researchers are limited to clearly grasp it. For this reason, there is a limit to apply standard sampling methods such as random sampling, systematic sampling, or stratified sampling for the entire population.

Furthermore, the survey measure is likely to cause self-report bias of respondents. This is because people inform their experiences or behaviors by themselves. In reality, according to the OECD Guidelines, self-report ways are possible to have various bias or heuristic such as extreme responding, nay-saying, random responding, etc due to social desirability or recall inaccuracies (OECD, 2013).This bias can be noticeable specially on sensitive topics such as gambling tendency or money spending. Therefore, participants' underreport or overreport can affect the accuracy or reliability of the data.

Lastly, this questionnaire of the research is altered in the context of Gacha games based on the PGSI. However, there might be a lack of verification of how these modifications ensured the validity of the PGSI. Due to this, it should be taken care when interpreting research results.

*4.5 Ethical Consideration*

Every participant could answer the survey anonymously to ensure privacy as well as increase involvement. The questionnaire asks only the information necessary for the research purpose, such as respondent's age to start playing Gacha games, their propensity in the games, etc., and does not collect any sensitive information that could be personally identifiable, such as name, current address, age, educational background, and family situation. This allows respondents to answer honestly without worrying about leaking their personal information.

In addition, every respondent is able to participate in the questionnaire voluntarily. In other words, people have the right to choose whether they engage in the survey or not, and there is no disadvantage if they do not answer it. This spontaneity creates the environment that makes participants willing to respond to it.

The survey clearly explains the objective and procedure of the study before people participate in it. By providing sufficient information on the intention of the study and the research team who conducts the survey at the beginning of the survey, respondents could understand the meaning of the research and answer it with confidence. Namely, this study has transparency, taking apart honestly.

Lastly, it can make sure that the collected data will not be used for any purpose other than the study. After the study is completed, all the data will be safely deleted. This measure can make respondents' information secure, which increases the reliability of the study.

*[5. Research Results & Analysis]*

**H1 Regression Results: PGSI = b * (entry age) + a**

| SUMMARY OUTPUT | | | | | | | | |
|---|---|---|---|---|---|---|---|---|
| *Regression Statistics* | | | | | | | | |
| Multiple R | 0.362901058 | | | | | | | |
| R Square | 0.131697178 | | | | | | | |
| Adjusted R Squ | 0.109989607 | | | | | | | |
| Standard Error | 0.450115532 | | | | | | | |
| Observations | 42 | | | | | | | |
| | | | | | | | | |
| ANOVA | | | | | | | | |
| | df | SS | MS | F | Significance F | | | |
| Regression | 1 | 1.22917366 | 1.22917366 | 6.066877797 | 0.018174403 | | | |
| Residual | 40 | 8.104159674 | 0.202603992 | | | | | |
| Total | 41 | 9.333333333 | | | | | | |
| | | | | | | | | |
| | Coefficients | Standard Error | t Stat | P-value | Lower 95% | Upper 95% | Lower 95.0% | Upper 95.0% |
| Intercept | 0.927057244 | 0.250853763 | 3.695608286 | 0.000656922 | 0.420062878 | 1.43405161 | 0.420062878 | 1.43405161 |
| X Variable 1 | -0.036510109 | 0.014822809 | -2.463103286 | 0.018174403 | -0.066468122 | -0.006552095 | -0.066468122 | -0.006552095 |

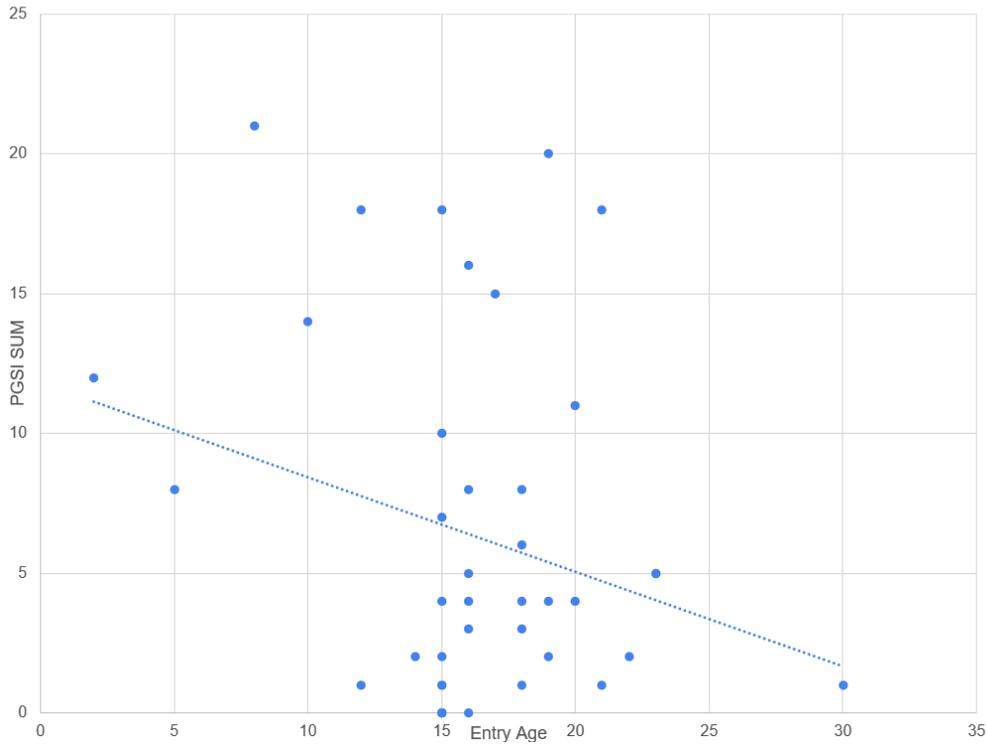

The linear regression results setting *x* value as the *sum of PGSI* and *y* value as the *entry age*. The most notable variable, *P-value* is *0.018 < 0.05*. Therefore, there is a significance between the *entry age* and the *sum of PGSI*. However, the coefficient value at *-0.03* is closer to *0* which indicates almost no correlation between the two variables.

On the contrary, when plotting a linear graph of the two variables, the direction and the slope of the graph does indicate there to be a weak connection between the two variables. Thus, there is a mismatch between the regression model and the linear graph model where the regression results show weak significance and the graph shows stronger significance.

**H2 Regression Results: PGSI = b \* (frequency (hours per day)) + a**

| SUMMARY OUTPUT | | | | | | | | |
|---|---|---|---|---|---|---|---|---|
| | | | | | | | | |
| Regression Statistics | | | | | | | | |
| Multiple R | 0.40642082 | | | | | | | |
| R Square | 0.16517788 | | | | | | | |
| Adjusted R Sc | 0.14377219 | | | | | | | |
| Standard Errc | 5.59607096 | | | | | | | |
| Observations | 41 | | | | | | | |
| | | | | | | | | |
| ANOVA | | | | | | | | |
| | df | SS | MS | F | Significance F | | | |
| Regression | 1 | 241.651211 | 241.651211 | 7.7165389 | 0.00836974 | | | |
| Residual | 39 | 1221.3244 | 31.3160102 | | | | | |
| Total | 40 | 1462.97561 | | | | | | |
| | | | | | | | | |
| | Coefficients | Standard Erro | t Stat | P-value | Lower 95% | Upper 95% | Lower 95.0% | Upper 95.0% |
| Intercept | 4.02702616 | 1.12065261 | 3.59346522 | 0.00090347 | 1.76029231 | 6.29376001 | 1.76029231 | 6.29376001 |
| 1 | 0.26845406 | 0.0966404 | 2.77786589 | 0.00836974 | 0.07298041 | 0.46392771 | 0.07298041 | 0.46392771 |

The linear regression results when plotting *x* value as the *sum of PGSI* and the *y* value as *duration(play time in hours per day)*, it shows more consistent results than hypothesis 1. The *P-value* is *0.008 < 0.05*. Thus, there is a significance between the *duration(play time in hours* *per day)* and the *sum of PGSI*. Here, the coefficient value is at *0.26* which indicates a weak to moderate relationship between the two variables.

The regression analysis results are further backed by the linear graph model. The graph shows a positive relationship between the *duration(play time in hours per day)* and the *sum of PGSI* which is inline with the coefficient value shown in the regression result.

## H3 T-test Results:

**Number of playing games )**

**F-Test: Two-Sample for Variances**

| | low-moderate | severe |
|---|---|---|
| Mean | 1.785714286 | 1.928571429 |
| Variance | 1.137566138 | 1.763736264 |
| Observations | 28 | 14 |
| Degrees of Freedom | 27 | 13 |
| F Ratio | 0.644975193 | |
| P(F<=f) One-Tail | 0.162839543 | |
| F Critical One-Tail | 0.475409336 | |

**t-Test: Two-Sample Assuming Equal Variances**

| | low-moderate | severe |
|---|---|---|
| Mean | 1.785714286 | 1.928571429 |
| Variance | 1.137566138 | 1.763736264 |
| Observations | 28 | 14 |
| Pooled Variance | 1.341071429 | |
| Hypothesized Mean Difference | 0 | |
| Degrees of Freedom | 40 | |
| t Stat | -0.376872452 | |
| P(T<=t) One-Tail | 0.354130195 | |
| t Critical One-Tail | 1.683851013 | |
| P(T<=t) Two-Tail | 0.708260391 | |
| t Critical Two-Tail | 2.02107539 | |

Variable: the number of Gacha games played

**Amount of money spent)**

**F-Test: Two-Sample for Variances**

| | low-moderate | severe |
|---|---|---|
| Mean | 93314.28571 | 408571.4286 |
| Variance | 11557610159 | 3.35798E+11 |
| Observations | 28 | 14 |
| Degrees of Freedom | 27 | 13 |
| F Ratio | 0.034418362 | |
| P(F<=f) One-Tail | 1.72E-12 | |
| F Critical One-Tail | 0.475409336 | |

**t-Test: Two-Sample Assuming Unequal Variances**

| | low-moderate | severe |
|---|---|---|
| Mean | 93314.28571 | 408571.4286 |
| Variance | 11557610159 | 3.35798E+11 |
| Observations | 28 | 14 |
| Hypothesized Mean Difference | 0 | |
| Degrees of Freedom | 13 | |
| t Stat | -2.01829615 | |
| P(T<=t) One-Tail | 0.032341606 | |
| t Critical One-Tail | 1.770933396 | |
| P(T<=t) Two-Tail | 0.064683212 | |
| t Critical Two-Tail | 2.160368656 | |

Variable: the amount of money spent

To test this hypothesis, a *T-test* involves two separate groups: *low-moderate* and *severe players* compared with the *number of playing gacha games* and the *amount of money spent on gacha games*.

First, to decide which kind of T-test had to be implemented (t-test assuming *unequal variance* or *equal variance*) before conducting T-test, *F-test* was conducted. For the *number of playing gacha games*, the P-value of the F-test came out as about 0.1628, which is bigger than the 0.05, meaning that T-test assuming equal variance is supposed to be done. According to the equal variance test, the *T-value* sits at about *-0.377* where the negativity indicates that the mean of the *severe players* is higher than the *low-moderate players* with a relatively small significance. Furthermore, the *P-value* is about *0.354> 0.05* thus, not being statistically significant.

For the *amount of money spent*, the P-value of the F-test was 1.72E - 12, which is an extremely small number, indicating smaller than 0.05. Thus, T-test assuming unequal variance had to be implemented. Based on the unequal variance test,  the *T-value* came out as *-2.018* and the negativity indicates the difference in terms of means as before. Again, the *P-value* resulted in about *0.032 > 0.05*. Thus, the test here proves that the relationships are significant.

## H1 & H2 Multiple Regression Results:

--- Start of R code

library(tidyverse)

df = read_csv("./data.csv")

# Fit a multiple regression model

model = lm(pgsi ~ entry_age + hours_per_week, data = df)

summary(model)

--- End of R code

[Analysis]

> summary(model)

Call:

lm(formula = pgsi ~ entry_age + hours_per_week, data = df)

Residuals:

  Min   1Q Median   3Q   Max

-7.997 -4.919 -1.214  2.420 15.785

Coefficients:

       Estimate Std. Error t value Pr(>|t|)

(Intercept)    9.2546   3.5423  2.613  0.0127 *

entry_age    -0.2768   0.1988 -1.392  0.1717

hours_per_week  0.2188   0.1037  2.111  0.0412 *

---

Signif. codes:  0 '***' 0.001 '**' 0.01 '*' 0.05 '.' 0.1 ' ' 1

Residual standard error: 5.971 on 39 degrees of freedom

 (100 observations deleted due to missingness)

Multiple R-squared:  0.1599,     Adjusted R-squared:  0.1169

F-statistic: 3.713 on 2 and 39 DF,  p-value: 0.03342

In the multiple regression analysis, the test was run through R with the *x* variable *sum of PGSI* and two *y* variables and *entry age* and *duration(hours per day)*. The results show that the *duration(hours per day)* is more significant than the *entry age* as the coefficient is *0.218* with *P-value 0.041 < 0.05* compared to the coefficient for *entry age -0.276* with *P-value 0.17 > 0.05* which is different from the linear regression results. Although it is a small improvement in the coefficient value, it is also worth noting that the coefficient of the *entry age* has improved compared to the linear regression results. The multiple regression analysis results support both linear regression results for hypothesis 1 and 2 while displaying clear differences in the significance of the two variables in terms of the *sum of PGSI* and that the relationship between *entry age* and the *sum of PGSI* is inconclusive.

*[6. Conclusion]*

As the gacha business model has become one of the most successful business models in the gaming industry, it has also affected many lives by dragging them to addiction.[33] This study's main concern was that mobile games with extremely effective design that are designated to hold people's attention and lead them towards addiction[34] has almost no age restriction. Thus, this research's main purpose was to find potential correlations between *entry age* and the *sum of PGSI* (indication of severity in gacha/gambling addiction) and spur discussion regarding this topic by setting up multiple hypotheses and a secondary research question.

The analysis of the survey results were analyzed using *linear regression analysis, T-test* and *multiple regression analysis.* Most of the results shown were either a weak to moderate relationship between the variables or an insignificant relationship. The *linear regression analysis* for hypothesis 1 showed that the correlation between *entry age* and the *sum of PGSI* was relatively insignificant with a negative linear graph when plotted directly into a linear graph model. However, test results for hypothesis 2 shows significance in the relationship between the *frequency(hours played per day)* and the *sum of PGSI.* These results are also verified by the *multiple regression analysis* where clear distinction in the coefficients and the significance levels are shown. In contrast to hypothesis 1 where some correlation could be spotted when plotted in a linear graph, test results for hypothesis 3 showed no significant relationship between *number of games played* but in terms of *amount of money spent* with the *sum of PGSI, there is statistical significance*.

The research findings highlight that while the entry age into gacha games showed a weak correlation with gambling severity, the frequency and duration of game play were significantly related to higher gambling severity. Specifically, players who spent more time per day engaged in gacha games tended to have higher gambling severity, as measured by the Problem Gambling Severity Index. This suggests that prolonged exposure and frequent engagement are more influential factors in developing gambling-like behaviors than the age of initial exposure alone. Moreover, the hypothesis that severe gambling behavior would lead to increased involvement was not supported by the data, but increased expenditure across multiple gacha games was supported, indicating that while severity correlates with increased spending, it does not necessarily predict the engagement with additional games. This finding points to the need for a nuanced understanding of how gambling behavior develops and the role of individual differences in susceptibility to gacha game addiction.

Given these results, several implications arise for policymakers, public health professionals, and educators. Although the correlations are between weak and moderate, there is a clear need for stricter regulation and oversight of mobile gacha games, particularly to protect younger and more vulnerable populations. The mobile gacha games themselves were proven that their designs were effective at getting adolescents to be addicted.[35] Therefore, the combination of wide availability of mobile phones, no age restrictions in mobile gacha games and the fact that adolescents are more vulnerable to gambling addiction[36], further actions from the gacha game companies, education system and the government must be taken.

Future research should aim to address the limitations identified in this study, including the need for longitudinal data and sampling a larger population to better understand the long-term effects of gacha game exposure with the help of a more complex analysis model.

Exploring the role of individual cognitive biases and psychological vulnerabilities in gambling behavior could provide deeper insights into the mechanisms underlying gacha game addiction. This study should be viewed as a first step into venturing a fairly new business model that has become popular amongst the industry. With the collected evidence and logical connections made from this research, severe consequences of the industry's practice could be looming around the corner.